\documentclass[12pt]{article}
\usepackage[totalheight = 23cm, totalwidth = 17cm]{geometry}
\usepackage{amssymb,amsmath,amsfonts,amsbsy,graphicx}
\usepackage{color}
\usepackage{psfrag}
\usepackage{cite}
\usepackage{stackrel}
\usepackage{pstool}

\newcommand\be{\begin{equation}}
\newcommand\ee{\end{equation}}
\newcommand\ba{\begin{eqnarray}}
\newcommand\ea{\end{eqnarray}}\newcommand\eq{\begin{equation}}           
\newcommand\en{\end{equation}}

 \newcount\colveccount
\newcommand*\colvec[1]{
        \global\colveccount#1
        \begin{pmatrix}
        \colvecnext
}
\def\colvecnext#1{
        #1
        \global\advance\colveccount-1
        \ifnum\colveccount>0
                \\
                \expandafter\colvecnext
        \else
                \end{pmatrix}
        \fi
}
\input epsf

\usepackage{ulem}
\usepackage{amssymb}
\usepackage{amsmath}
\usepackage{bm}
\usepackage{graphicx}
\usepackage{color}
\usepackage{subfig}
\usepackage{caption}

\def\gsim{\;\rlap{\lower 2.5pt
 \hbox{$\sim$}}\raise 1.5pt\hbox{$>$}\;}
\def\lsim{\;\rlap{\lower 2.5pt
 \hbox{$\sim$}}\raise 1.5pt\hbox{$<$}\;}

\usepackage{graphicx}
\usepackage{url}
\usepackage{amssymb,amsmath}
\usepackage{amsbsy}

\usepackage{graphicx}
\usepackage{url}
\usepackage{amssymb,amsmath}
\usepackage{amsbsy}

\begin{document}
\title{
  Cross-correlation between 21-cm radiation and CMB B modes from the cosmic birefringence in the presence of a light scalar field}
\author{Kenji Kadota$^1$, Junpei Ooba$^2$, Hiroyuki Tashiro$^2$, Kiyotomo Ichiki$^{2,3}$, Guo-Chin Liu$^4$ \\
 {\small  $^1$ \it Center for Theoretical Physics of the Universe, Institute for Basic Science (IBS), Daejeon, 34051, Korea}\\
  {\small $^2$
 \it Department of physics and astrophysics, Nagoya University, Nagoya 464-8602, Japan}\\
  {\small $^3$ \it Kobayashi Maskawa Institute, Nagoya University, Aichi 464-8602, Japan}
  \\
 {\small $^4$ \it Department of Physics, Tamkang University, Tamsui, New Taipei City 25137, Taiwan}
}
\date{\vspace{-5ex}}
\maketitle   

\begin{abstract}
  We study the cross-correlation between the 21cm and CMB B mode fluctuations which are induced by the cosmic birefringence when the non-constant scalar field couples to the electromagnetic field strength. 
  Such multi-wavelength signals can potentially probe the reionization history of the Universe and also explore the nature of fundamental physics such as the parity violation and the scalar field dynamics in the early Universe.
  We illustrate the feasibility to detect such 21cm-B mode cross-correlations through commonly discussed scalar field models, the quintessence-like scalar field which is responsible for the current dark energy and the axion-like scalar field which is responsible for the current dark matter density.
\end{abstract}

\setcounter{footnote}{0} 
\setcounter{page}{1}\setcounter{section}{0} \setcounter{subsection}{0}
\setcounter{subsubsection}{0}

\section{Introduction}

We study the birefringence effects due to the parity violating nature of the pseudo-scalar $F _{\mu\nu}\tilde{F}^{\mu\nu}$ ($F$ is the electromagnetic field strength) when it couples to the (pseudo-)scalar field $\phi$.
In existence of such a term $\phi F _{\mu\nu}\tilde{F}^{\mu\nu}$, the polarization vector of a photon is rotated by an angle corresponding to the change in $\phi$ as the photon propagates. 
Even if its coupling is small, a total rotation can be appreciable thanks to a large cosmological distance traveled by a photon (referred to as cosmic birefringence).  We study the potential signals of the CMB B modes which are induced from the E modes by the cosmic birefringence, in particular their cross-correlation with the 21cm signals. Many searches for the cosmological birefringence effect as a promising astrophysical test of the fundamental physics have been attempted, such as the CMB and radio galaxy polarization observations \cite{car1989, car1991,hara1992,car1997,lue1998,feng2004,liu2006,liu2016,fe2006,kamion2008,posp2008, li2009,cal2011, yada2012,ade2015cao, ag2016b,arr2017,sigl2018fba,fedd2019,pog2019}. For instance, the parity violating interaction term can lead to the parity-odd combinations for CMB cross-correlations $C_l^{TB}, C_l^{EB}$ which would otherwise vanish due to the parity conservation. 
Cross-correlating the small B mode signal with a much larger CMB temperature signal can potentially make such birefringence-induced B modes more favorable for detection, similar to the situation where $C_l^{ET}$ is bigger than the E mode auto-correlation power spectrum.
With the 21cm brightness temperature fluctuations even bigger than the CMB temperature fluctuations, studying the multi-wavelength cross-correlation between the birefringence-induced B mode and the 21cm signals would be worth exploring which can also benefit from the different noise dependence compared with the auto-correlations. 

   The 21 cm line emissions from the neutral hydrogen can provide us a unique clue on the properties of the high redshift Universe such as those during the the dark ages and epoch of reionization \cite{fur2006}. The recent 21cm global temperature results, for instance, drew much attention from the particle physics community as well for the dark matter study \cite{bow2018,tashi2014,muno2015,bark2018,ber2018,fra2018,kov2018,siki2018}.  The fluctuation of the 21 cm temperature which is the focus of this paper is also of great interest for the precision cosmology and many 21cm experiments are in progress and planned \cite{MWA, PAPER, LOFAR, HERA, SKA}.
   Extracting the information from those radio frequency maps are however plagued with the astrophysical and instrumental noises, and there have been active studies for cross-correlating the 21cm signals with other observables such as the CMB and high redshift galaxies which can provide the additional information besides the auto-correlation statistics alone as well as improving the distinction from the foregrounds/systematics. For instance the CMB photons are scattered by the ionized electrons and the cross-correlation between the CMB and the 21cm can arise, and many noises and foregrounds which the low frequency radio observations suffer from can be more controllable or uncorrelated with the CMB measurements \cite{ala2005,ad2007,coo2004,coor2007,tashib2008,tashi2009uj,sal2005,liz2008c,kubo2017}. There have been related works which discussed the correlations between the 21 cm and CMB temperature/E-mode anisotropies from the reionization epoch at large scales \cite{ala2005,ad2007,tashib2008,tashi2009uj}. The large linear scale has an advantage where the linear treatment of the fluctuations is applicable and the complicated patchy reionization effects are averaged out \cite{sal2005,coo2004}.

   Motivated by those active studies on the cross-correlation measurements, we study the cross-correlation of the 21 cm signals with the cosmic birefringence signals from the reionization epoch at the linear scale. More specifically, we discuss the cross-correlation between the 21cm and birefringence-induced CMB B mode in existence of the background scalar field, which can potentially give us an additional probe on the the scalar field evolution in the early Universe as well as the nature of reionization epoch. Even though the physics gain through the detection of 21cm-B mode cross-correlations is attractive, we illustrate that it is challenging and, at least for our choice of simple reionization and scalar field models, unlikely to detect such a cross-correlation signal with the forthcoming experiments. We however also illustrate that the signals could be enhanced/reduced significantly depending on the reionization history and the scalar field model, so that 21cm-B mode cross-correlation signals, if detected, can potentially take advantage of such a strong model dependence to elucidate the properties of the background scalar field and the reionization process.


We outline the formalism to calculate the 21cm and B mode fluctuations in Section \ref{sectioncross}. The numerical results for the cross-correlation calculations are presented for the concrete models, the quintessence-like dark energy scalar and the axion-like ultra-light dark matter models in Sections \ref{results} and \ref{discon}.


\section{The 21cm-B mode cross-correlations}
\label{sectioncross}
We first outline the setups to calculate the cross-correlations between the 21 cm and B mode fluctuations in this section, before presenting the numerical results using the concrete scalar field models in the following sections.

 The birefringence-induced B mode arises from the E mode due to the polarization plane rotation. The CMB E mode is generated when the CMB temperature anisotropy quadrupole scatters off free electrons, notably at the recombination ($z\sim 1100$) and the reionization epochs (e.g. $z\sim 10$) \cite{hu1997hv}.
The reionization epoch in particular is among the main targets of the 21cm observations and the 21cm signals are expected to unveil the nature of reionization which still remains an open question in the history of the Universe \cite{fur2006}.  
We hence focus on the reionization epoch in the following analysis on the 21cm-B mode cross-correlation because this is the common epoch  which both 21cm and B mode fluctuations are susceptible to and hence appreciable cross-correlations can potentially arise.

\subsection{Cosmic birefringence}

We consider the term where the scalar field $\phi$ couples to the electromagnetic field strength
\ba
L\ni -\frac{\beta}{4} \phi F_{\mu \nu}\tilde{F}^{\mu \nu}
\ea
 which modifies the Maxwell equations and, due to its parity violation nature of the pseudo scalar $ F_{\mu \nu}\tilde{F}^{\mu \nu}$, the left and right circularly polarized photons have different dispersion relations resulting in the rotational speed of the linear polarization plane in the direction $\hat{n}$ ($\eta$ is the conformal time) \cite{car1989, car1991, hara1992}
\ba\omega(\vec{x},\eta)=-{\beta\over2}\left(\frac{\partial \phi}{\partial\eta}+
\vec{\nabla} \phi \cdot \hat{n}\right)
\ea
If $\phi$ is constant in space and time, such a term is merely a total derivative and has no effect on electrodynamics. We study the scenario where the photon propagates in the homogeneous background of $\phi(t)$ which varies with time. The rotation angle from the emitted epoch $\eta$ to the current conformal time $\eta_0$ is
\ba
\label{rotan}
\theta(\eta)=\int _{\eta}^{\eta_0}d\eta \omega
\ea
When the field is homogeneous 
\ba
\theta(\eta)=  - \int _{\eta}^{\eta_0}d\eta
\frac{\beta}{2} \frac{\partial \phi}{\partial \eta}
=-\frac{\beta}{2} (\phi(\eta_0)-\phi(\eta))
\ea

Note this rotation is independent of the frequency and hence can be distinguished from the frequency dependent Faraday rotation by the multi-frequency maps \cite{koso1996}. Such a rotation of the polarization plane mixes the spin-2 Stokes parameters as $(Q(\hat{n}) \pm i U(\hat{n}))
\rightarrow e^{\mp 2 i \theta }
\left[
  Q(\hat{n}) \pm i U(\hat{n})
  \right]$ and consequently mixes the E and B modes which can be expressed in terms of the Stokes parameters \cite{sel1996,zald1996,hu1997}\footnote{We consider the CMB polarization which is produced from the Thomson scattering, and the Stokes parameter $V$ representing the net circular polarization remains zero if it is zero initially even in existence of $\phi F \tilde{F}$ term \cite{li2008}.}. We for simplicity focus in this paper only on the the B modes which are converted from the E modes due to the birefringence at low $l\lesssim 100$ where the linear theory is applicable \footnote{Other sources of B modes include, for instance, those from gravitational waves, gravitational lensing and patchy reionization, which would be also worth exploring.}.

\subsection{21 cm fluctuations}
The observed differential brightness temperature of the 21 cm line emitted at a redshift $z$ is 
\cite{field59, madau1996}
\ba
\label{tb21}
T_{21}=
\hat{T}_{21}
x_{HI}
(1+\delta_b)
\left(
1-\frac{T_{CMB}}{T_S}
\right),
\hat{T}_{21}=
23[\mbox{mK}]
\left(
\frac{\Omega_b h^2}{0.02}
\right)
\left[
  \left(
  \frac{0.15}{\Omega_m h^2}
  \right)
  \left(
  \frac{1+z}{10}
  \right)
  \right]^{1/2}
\ea
$x_{HI}$ is the neutral hydrogen fraction, and $\delta_b$ is the baryon density contrast.
In terms of $\delta_b$ and the neutral fraction fluctuation $\delta_{HI}\equiv x_{HI}/\bar{x}_{HI}-1$, the brightness temperature fluctuation up to the linear order is $ T_{21}=\hat{T}_{21} \bar{x}_{HI} (\delta_{HI} + \delta_b)$ (the overbar denotes the average). We focus on the redshift regime where the spin temperature exceeds the CMB photon temperature $T_S\gg T_{CMB}$ to simplify our analysis, which is a valid assumption soon after the reionization starts \cite{cia2003}. 
The baryon density contrast $\delta_b$ has already caught up with the matter density contrast $\delta_m$ for the redshift range of our interest $\delta_b \approx \delta_m$. 
For the ionization fraction fluctuations $\delta_{HI}$, we follow the model of Ref. \cite{ala2005}, $\delta_{HI}\approx  (\bar{b}-1-\epsilon) \ln \bar{x}_{HI}    \delta_m$, based on the conventional halo models with the averaged halo bias $\bar{b}$ 
\ba
\bar{b}=\frac{\int^{\infty}_{\nu_{\min}}    d \nu f(\nu) b(\nu)  } {\int^{\infty}_{\nu_{\min}} d\nu f(\nu) }
\ea
where $f(\nu)\propto \exp(-\nu^2/2)$, $\nu\equiv \delta_c/\sigma(M)$ is the threshold in units of the rms density fluctuations and $b$ represents the halo bias of Ref. \cite{mo1995} according to the Press-Schechter formalism.
$\nu_{\min}$ corresponds to the minimal halo mass with the temperature $T_{min}(\sim 10^4$K) above which the atomic hydrogen line cooling becomes efficient to induce the ionizing sources \cite{bar2000}.
 $\epsilon$ represents the nature of ionization process, and $\epsilon=1$ corresponds to the limit where the emitted photons to ionize the gas are balanced by the recombination and $\epsilon=0$ is for the case where the recombination is negligible. $\bar{x}_{HI} \rightarrow 1$ corresponds to the epoch when all emitted photons ionize the atoms before HII regions start percolating while $\bar{x}_{HI} \rightarrow 0$ corresponds to the epoch with much more ionizing photons than the atoms as expected after the HII regions percolate. We neglect the effects of the recombination in our numerical calculations for simplicity. Even though our simple halo model treatment would be reasonable (because the ionizing sources are formed in the denser regions and we are interested in the scales larger than a typical ionized bubble size), the actual ionization fluctuation estimate should be more complicated and heavily dependent on the reionization model and intergalactic medium (IGM) gas dynamics.
 This simple estimation would however suffice for our purpose of studying the future prospect of the cross-correlation between the CMB B mode and 21cm fluctuations and more detailed treatments are left for future work.


\subsection{The cross-correlations between 21cm and CMB B mode fluctuations}
We numerically obtained, by the modified version of CLASS \cite{class2011a}, the cross-correlation between the birefringence-induced B mode and 21 cm fluctuations
\ba
C_l^{B21}
=
\frac{2}{\pi} \int _0^{\infty} k^2 dk 
 P(k)
\Delta^{B}_l(k,z)\Delta^{21}_l(k,z)
\label{crossfor}
\ea
where $P(k)$ is the power spectrum of the primordial curvature perturbation and $\Delta^B_l, \Delta^{21}_l$ represent the transfer functions, respectively, for the B mode polarization and 21cm fluctuations. While our 21cm transfer function is the conventional one without being affected by the birefringence \cite{fur2006, madau1996, zalb2003, mc2005, tashib2008}, $\Delta^B_l$ for the B mode induced by the cosmological birefringence differs from the conventional B mode transfer function and resembles that for the E mode with an additional rotation angle factor $\sin [2 \theta(\eta)]$ \cite{hu1997,lestra2013}
\ba
\label{btransfer}
\Delta^B_l(k)=\sqrt{\frac{3(l+2)!}{8(l-2)!}} \int^{\eta_0}_{0} d\eta g(\eta) S_P(k,\eta)\frac{j_l(x)}{x^2} \sin [2\theta(\eta)]
\ea
  where $g$ is the visibility function, $j_l$ is the spherical Bessel function, $x=k(\eta_0-\eta)$ and $S_P$ is the source function (we consider only the dominant scalar perturbation source term)\cite{hu1997,lestra2013,sel1996b,hub1997}. 

  \section{Results}
  \label{results}
\begin{figure}
    \includegraphics[width=1.0\textwidth]{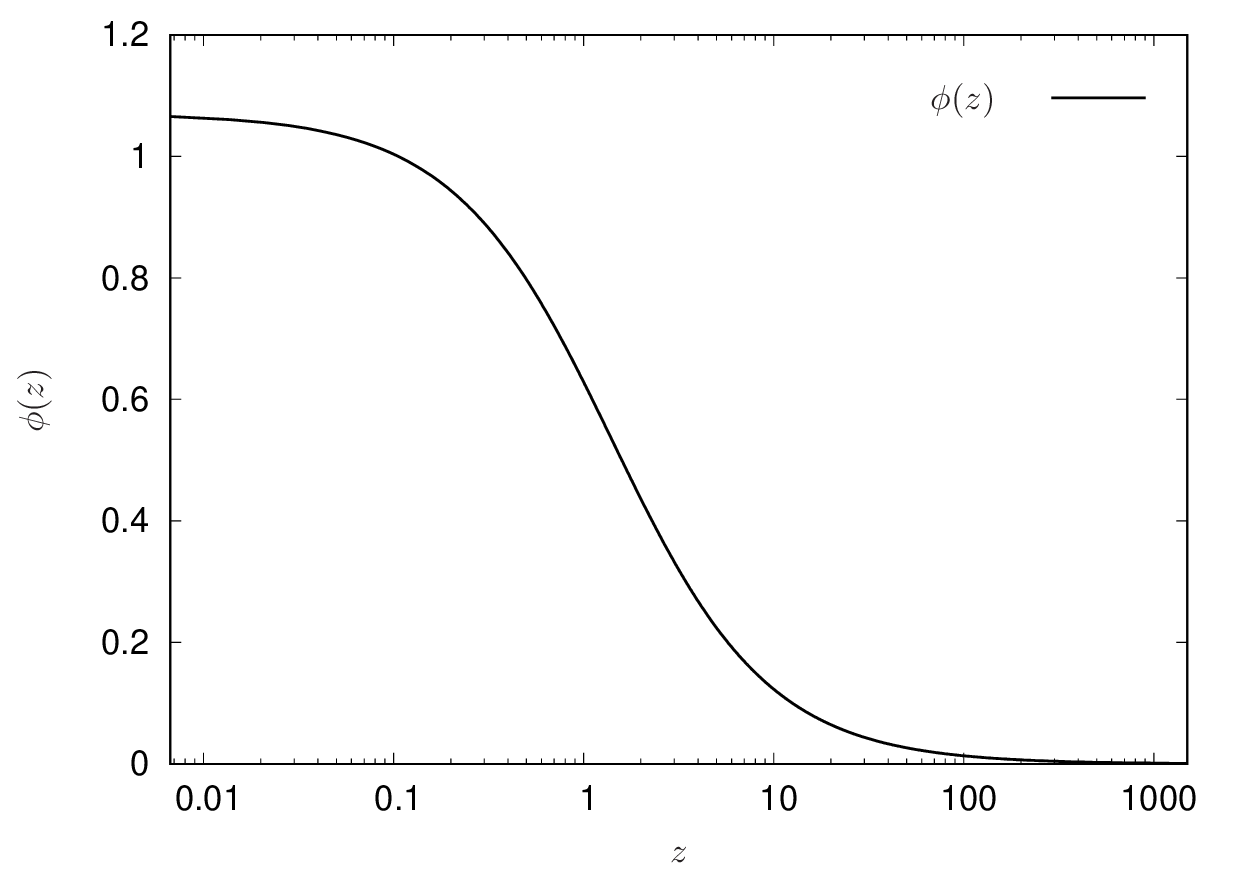}
        \caption{The dark energy scalar field evolution (in units of the reduced Planck scale).}
  \label{xemay14figleft}
\end{figure}

  We quantitatively illustrate the cross-correlations with a concrete example in this section.
For the numerical analysis, we need to specify the background scalar field evolution.
As a toy example, we introduce a simple scalar field dark energy model with the inverse power low potential 
\ba
V(\phi)=\Lambda^{4+\alpha} \phi^{-\alpha}
\ea
which possesses the tracker solution for the dynamical dark energy (Ratra-Peebles quintessence model) \cite{rat1987,pee1987,pee2002,pav2013,ooba2017}. We for concreteness  set $\alpha=1$ in our numerical analysis and a constant $\Lambda$ is numerically adjusted to match the observed dark energy density assuming the flat cosmology \cite{agh2018a}. The time evolution of $\phi$ is given in Fig. \ref{xemay14figleft} (we use in this paper the convention $M_p\equiv 1/\sqrt{8\pi G}=1$ unless specified otherwise) from which we can calculate the time dependent rotation angle and consequently the birefringence-induced B mode fluctuations. 

  The current CMB measurements put the bounds on the birefringence rotation angle of order $\Delta \theta\lesssim 0.5^{\circ}$ \cite{ag2016b,arr2017, sigl2018fba}.
We for the illustration purpose choose $\beta=0.01$ as a reference value which gives the total rotation angle of order $\Delta \theta =\beta \Delta \phi /2\sim 0.3 ^{\circ}$ ($\Delta \phi$ is the total field displacement) \footnote{Note the CMB bounds as well as many literature on this subject use the constant rotation angle approximation, which in some models could over/under-estimate the birefringence effects compared with more precise time dependent rotation angle analysis as is numerically performed in our study \cite{fine2008,gub2014,liu2006}.}.

\begin{figure}
    \includegraphics[width=1.0\textwidth]{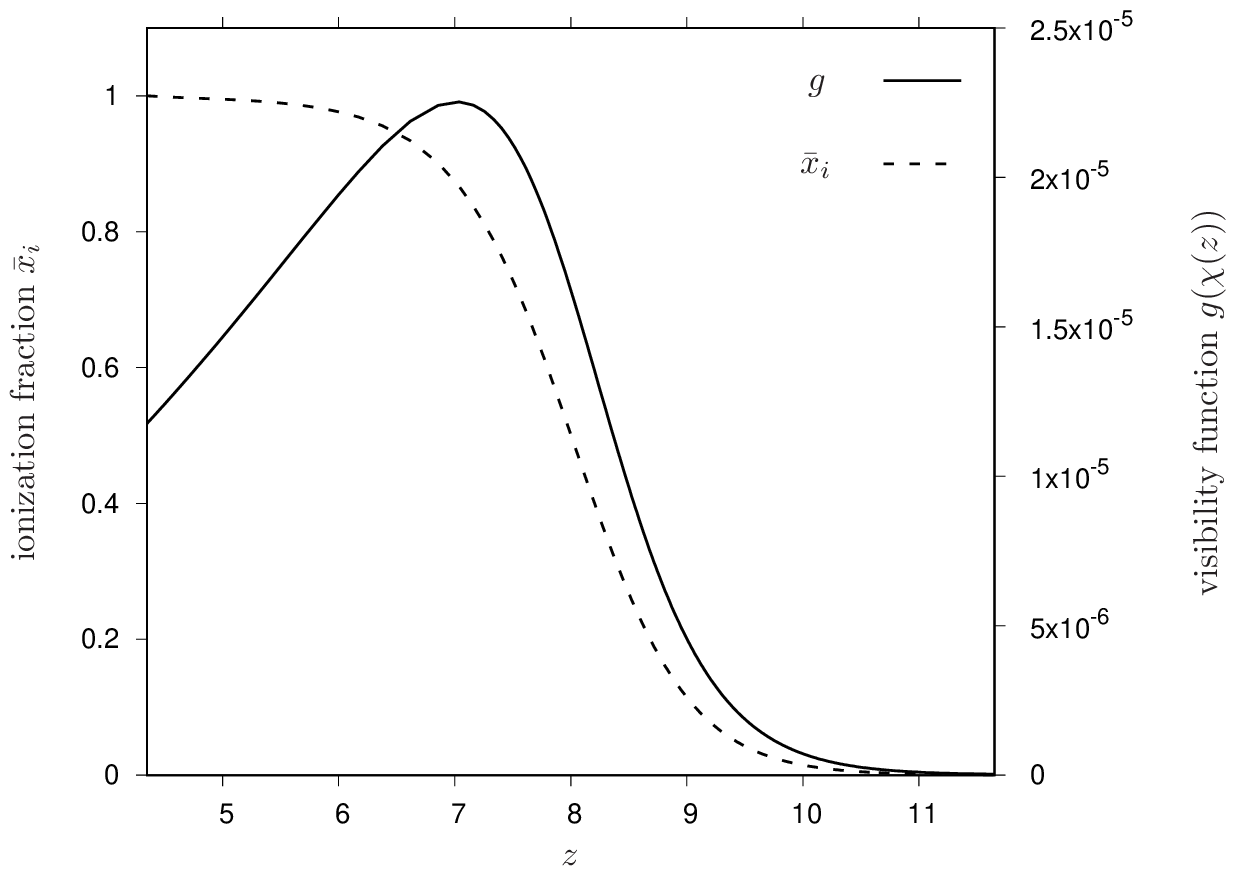}
       \caption{The ionization fraction and the visibility function.}
  \label{xemay14figright}
\end{figure}

To estimate the 21cm signals, one also needs to specify the reionization model. We use a simple model where the ionization fraction is parameterized by a tanh step $\bar{x}_i(z) =(1+ \tanh [(y(z_{rei})-y(z))/\Delta y ])/2$ with $y(z)=(1+z)^{1.5}, \Delta y=1.5 (1+z)^{0.5} \Delta z_{rei}$. This model has two free parameters, $z_{rei}$ and $\Delta z_{rei}$, and leads to the transition centered at $z=z_{rei}$ (when the ionization fraction is half) with the width represented by $\Delta z_{rei}$. 
 In the following numerical analysis, we simply set $\Delta z_{rei}=1$ and $z_{rei}$ was obtained numerically by matching the optical depth to the Planck value of $\tau=0.056$ \cite{agh2018a}.
The other parameters are also set to the flat $\Lambda$CDM cosmology from Planck \cite{agh2018a}. The evolution of ionization fraction $\bar{x}_i$ for such a choice of parameters is shown in Fig. \ref{xemay14figright} along with the visibility function which also depends on the reionization model. For our parameter set, this figure shows $\bar{x}_i=0.5$ at $z\sim 8$ and we simply call this model the $z_{rei}=8$ scenario for brevity.


     \begin{figure}

           \includegraphics[width=1.0 \textwidth]{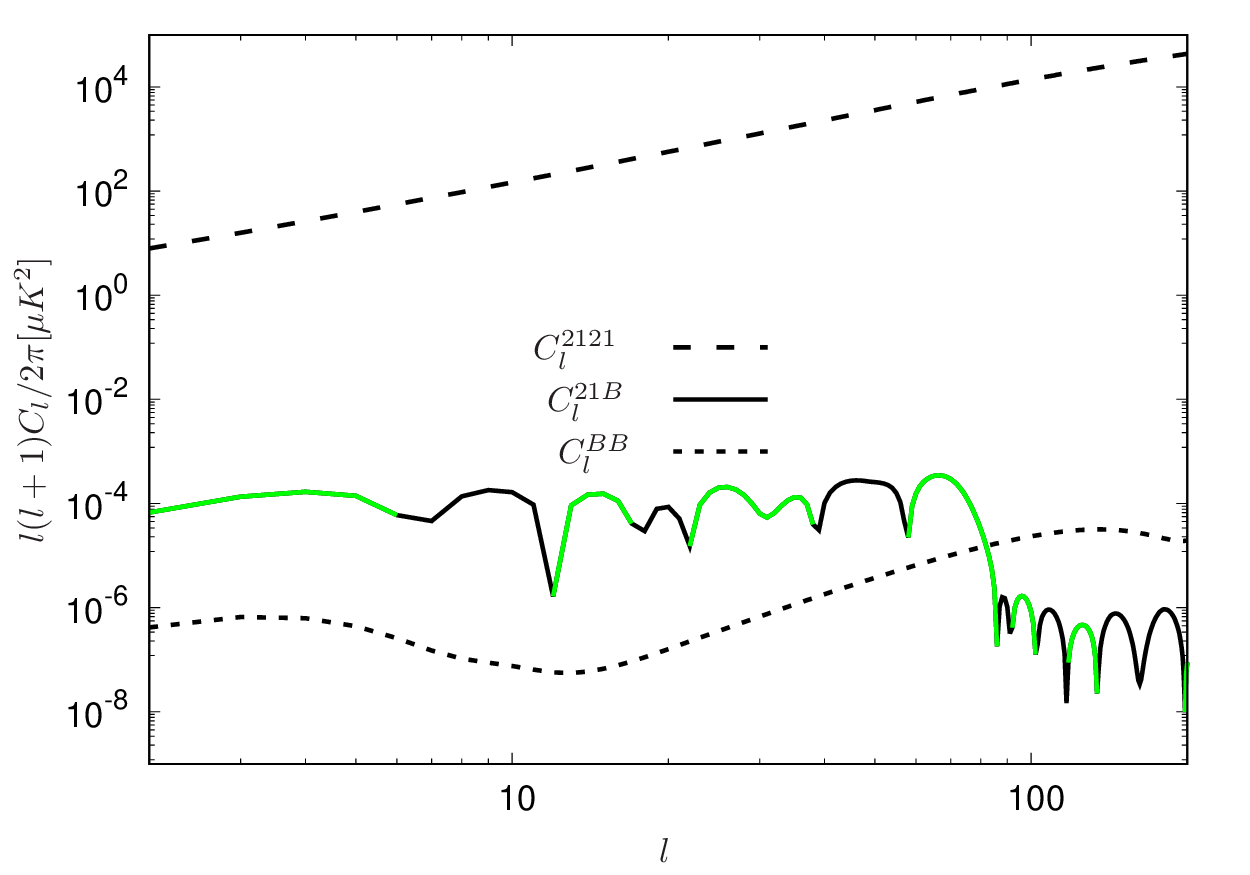}

  \caption{The power spectra $C_l$ for $\beta=0.01, z_{obs}=8$. For the cross correlation $C_l^{21B}$, the absolute values are plotted for the log plot (the negative (positive) correlations are indicated in green (black)). 
  }
  \label{cellsignalsleft}
  \end{figure}


            \begin{figure}
  
               \includegraphics[width=1.0 \textwidth]{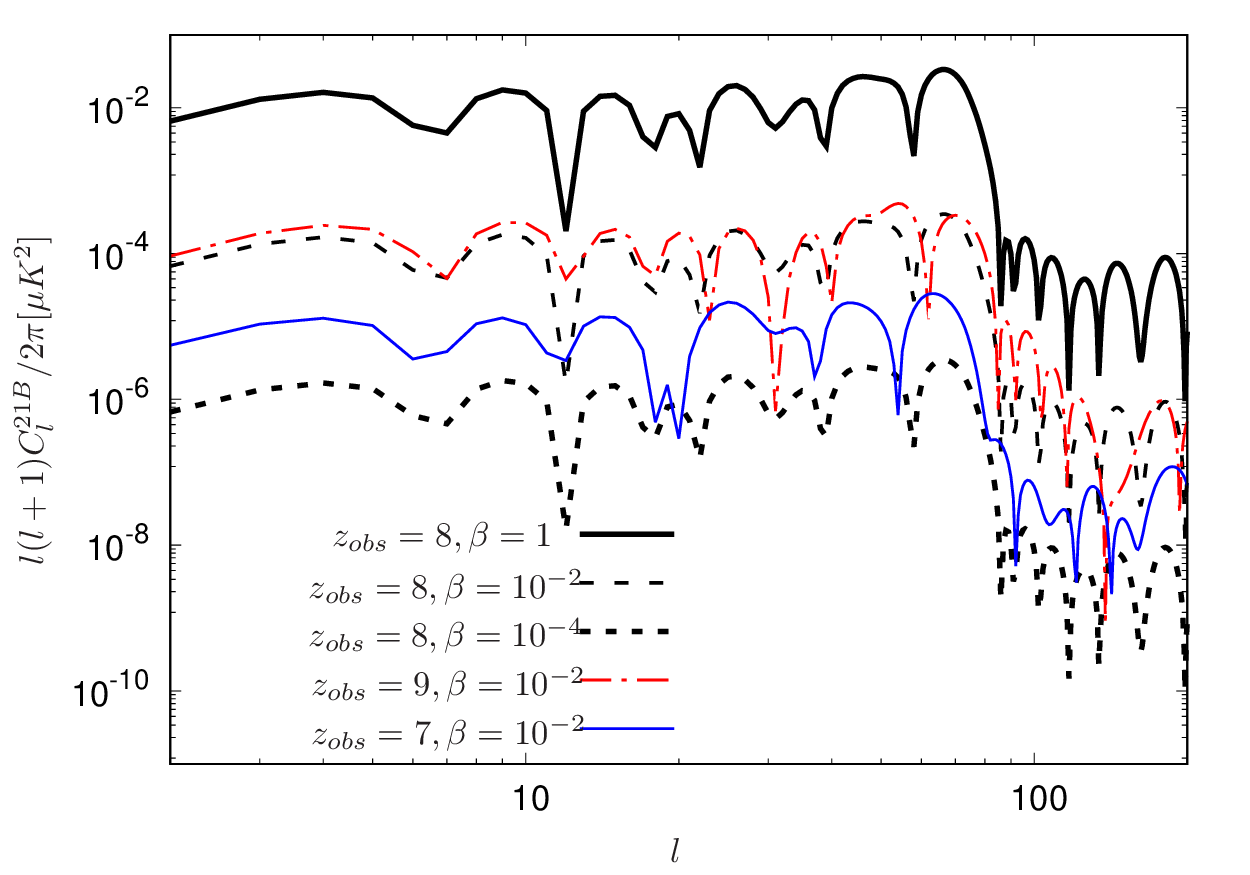}
  \caption{The cross-correlation $C_l^{21B}$ with $\beta=1,10^{-2},10^{-4}$ for $z_{obs}=8$. $C_l^{21B}$ with $\beta=0.01$ for $z_{obs}=7,9$ are also shown. 
  }
  \label{cellsignalsright}
            \end{figure}

The 21cm-B mode cross-correlation as well as the 21cm and B mode auto-correlation angular power spectra for the observation redshift $z_{obs}=8$ (corresponding to the observed 21cm line wavelength $\lambda_{obs}=21(1+z_{obs})$cm) is shown in Fig. \ref{cellsignalsleft}. In this log plot, we plotted the absolute values for the cross-correlation and indicated the negative correlation with the green color and the positive one with black. In the rest of the papers, we simply plot the absolute values for the cross-correlations in the log plot in our discussions. The B-mode auto-correlation $C_l^{BB}$ has two peaks in this figure. The peak at a higher $l$ is sourced form the recombination epoch and the other one at a lower $l\lesssim 10$ is sourced from the reionization epoch. The latter originates from the reionization bump of the E mode due to the birefringence and the shape of the birefringence-induced B mode power spectrum indeed resembles that of the E mode. Even if the amplitude of the B mode auto-correlation $C^{BB}_l$ is small, the cross-correlation $C_l^{21B}$ amplitude can be much bigger than $C_l^{BB}$ because of the large amplitude of 21cm fluctuations, even though the cross-correlation amplitude is not as large as the simple product $\sqrt{C_l^{2121}C_l^{BB}}$ because the correlation is not expected to be perfect. The 21cm angular power spectrum $C_l^{2121}$ is not affected by the cosmic birefringence. 
The peak amplitudes for the oscillations of $l(l+1) C_l^{21B}$ dot not change significantly at a low $l\lesssim 100 $ partly because $l(l+1) C_l^{2121}$ is bigger for a higher $l$ and this increase can compensate the decrease of the B mode.

The first peak of the $C_l^{21B}$ corresponds to the quadrupole (which sources the polarization) at the reionization epoch and can give us the information on when the reionization occurred \cite{tashib2008,tashi2009uj}. The oscillatory behavior of $C_l^{21B}$ arises because the quadrupole anisotropy at a higher redshift shows up at a higher multipole by the free-streaming effects.
%
A further suppression of the amplitude at a smaller scale $l \gtrsim 100$ is due to the finite width effects (the anisotropies are damped due to the cancellation among positive and negative fluctuations on scales smaller than the reionization width), which can give us the information on the reionization duration. The 21cm-B mode cross-correlation signals hence can potentially provide us with the crucial information to study the epoch of reionization.

Fig. \ref{cellsignalsright} also shows, in addition to the reference parameter values of $\beta=10^{-2},z_{obs}=8$, the plots for different choices of $\beta$ and $ z_{obs}$ (using our reference reionization model with $z_{rei}=8,\Delta z_{rei} =1$). The different behaviors for $\beta=1,10^{-2},10^{-4}$ are simply because the rotation angle scales proportionally to $\beta$ and hence the signal also scales in the same manner.  We also showed the cross-correlation for $z_{obs}=7,8,9$ for $\beta=10^{-2}$. The redshift dependence is more involved and cannot be described by a simple scaling. The redshift dependence for instance can come from the bias (which can become larger for a higher z) and the neutral hydrogen fraction (which can vary from 1 to 0 depending on a redshift). The redshift dependence also arises due to the visibility function which can peak during the reionization epoch. The signal for $z_{obs}=9$ benefits from the larger bias but suffers from the smaller visibility function, and that from $z_{obs}=7$ suffers from the smaller bias and smaller neutral hydrogen fraction compared with the signal from $z_{obs}=8$. The redshift dependence of the signal can also arise from the redshift dependence of the background scalar evolution.  Even though we cannot illustrate a simple redshift dependence of the cross-correlation signals, we can infer from this figure that it would not be trivial to obtain a significant signal enhancement (say by well more than an order of magnitude) compared with the signal when the reionization fraction reaches a half ($z_{obs} \sim 8$ for the toy model under discussion) by changing the observation redshift.  
For the following estimations of the signal to noise ratio (S/N), we simply consider $z_{obs}=8$ as a reference value.

Having estimated the cross-correlation signals, let us now discuss the feasibility for its detection. We estimate the 1-$\sigma$ error of the cross-correlation as \cite{knox}
\ba
\label{errorbar}
\Delta C_l^2
=\frac{1}
{(2l+1)f_{sky}\Delta l}
\left[
  \left(
C_l^{21B}
  \right)^2
  +
  \left(
  C_l^{BB}+N_l^{BB}
  \right)
  \left(
  C_l^{2121}+N_l^{2121}
  \right)
  \right]
\ea
and the signal to noise ratio accordingly can be estimated as
\ba
\label{snformula}
\left(
\frac{S}{N}
\right)^2
=
\sum_l
(2l+1) f_{sky}
\frac{\left( C_l^{21B} \right)^2}{ \left( C_l^{21B} \right)^2 +  \left(
  C_l^{2121}+N_l^{2121}
  \right)
  \left(
  C_l^{BB}+N_l^{BB}
  \right)   }
\ea
$\Delta l$ is the binning size and $f_{sky}$ is the sky fraction of the observation.
We estimate the noise power spectra for the 21cm and B modes as \cite{zalb2003,kata2011}
\ba
N_l^{21 21}=\frac{1}{2\pi t_{obs} \Delta \nu } \left( \frac{\lambda^2 l_{max}}{A/T} \right)^2 [\mu K^2 ~\mbox{str}]
\ea
and
\ba
N_l^{BB} =
\left(
\frac{\pi}{10800}
\frac{\omega_p^{-1/2}}{\mu K~ \mbox{arcmin}}
\right)^2
[\mu K^2 ~\mbox{str}]
\ea
$t_{obs}$ is a total observation time, $\Delta \nu$ is the bandwidth, $A/T$ (the effective area/system temperature) represents the telescope sensitivity and $l_{max}=2\pi  D/\lambda$ ($D$ is the baseline length and $\lambda=21(1+z)$ cm).
$\omega_p$ represents the polarization noise. In our numerical estimate, we included the effect of beam smearing by the Gaussian window function of  $\theta_{FWHM}=30$ arcmin $N_l^{BB} e^{l^2 \theta^2_{FWHM}/(8\ln 2)}$. 




\begin{figure}
                    \includegraphics[width=1.0\textwidth]{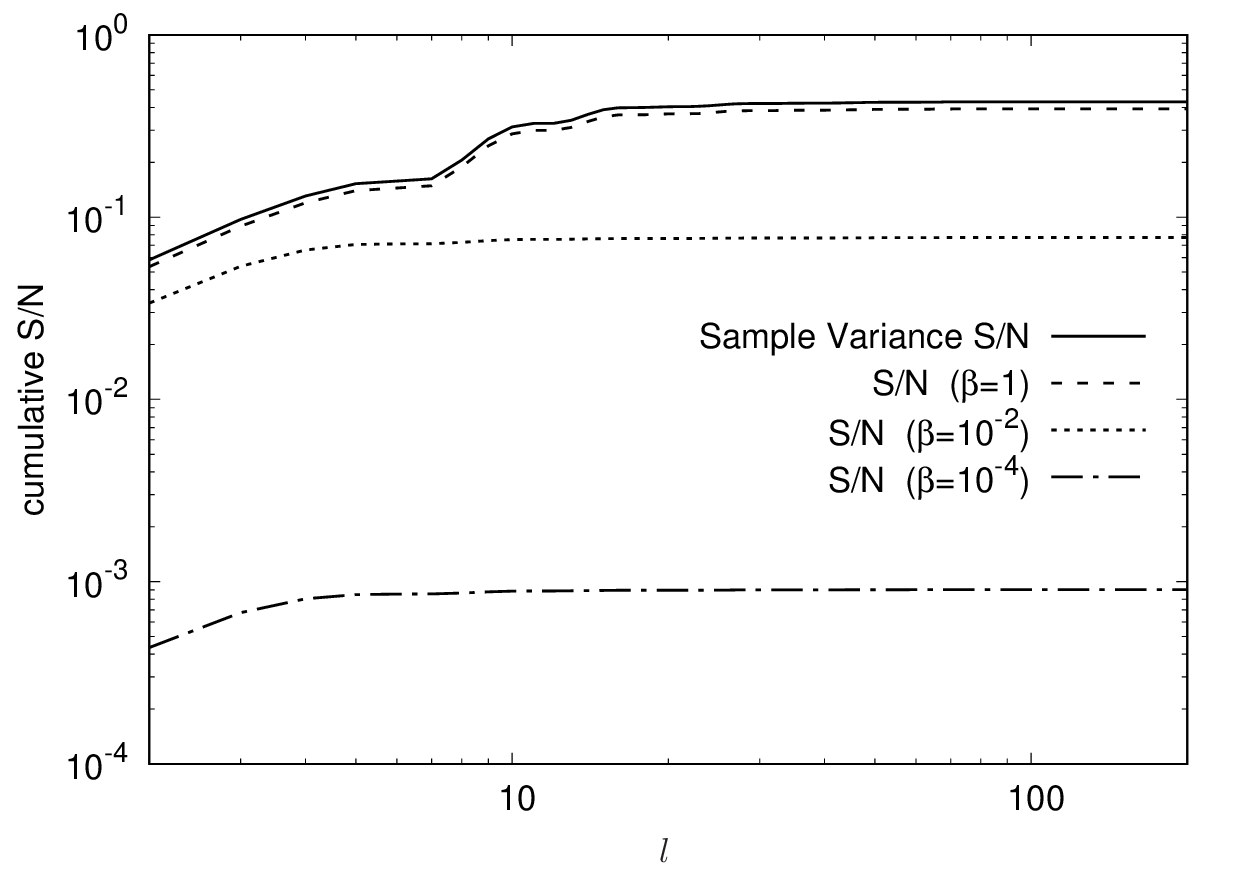}
        \caption{The cumulative signal to noise ratios (S/N) for the reionization model with $z_{rei}= 8$, $\beta=1,10^{-2},10^{-4}$. The sample variance limited S/N is also shown. 
        }
        \label{snratioplotleft}
\end{figure}

\begin{figure}
            \includegraphics[width=1.0\textwidth]{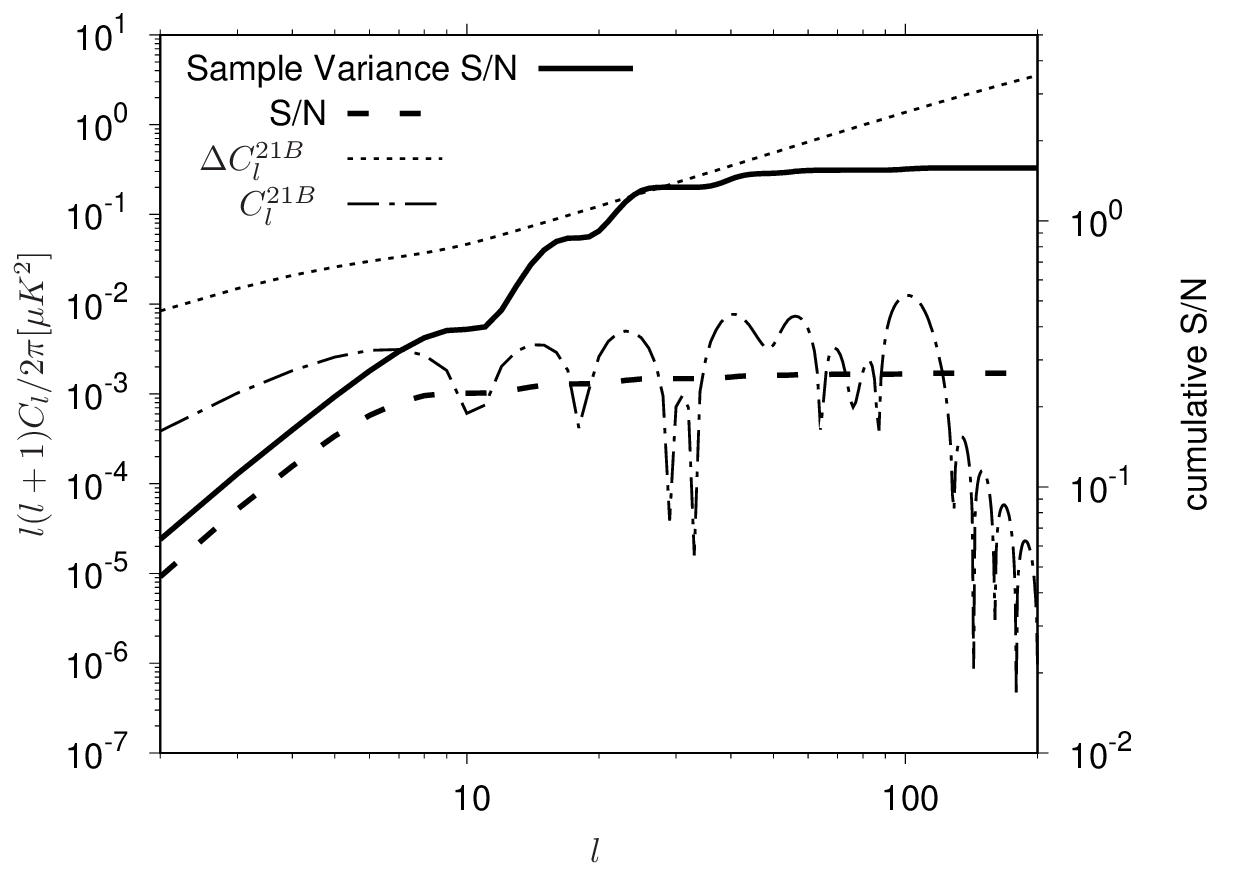}
        \caption{The signal $C_l$ and the error $\Delta C_l$ (Eq. \ref{errorbar}) are shown along with the cumulative signal to noise ratios for $z_{rei}=15,\beta=10^{-2}$.
  }
        \label{snratioplotright}
\end{figure}

The cumulative signal to noise ratios up to a given $l$ are shown in Fig. \ref{snratioplotleft}, where, for a concrete (optimistic) estimate, we used $f_{sky}$=1/2, $t_{obs}=1000$ hours, $\Delta \nu=1$MHz, $D=1km,A/T=2000 m^2/K$ \cite{SKA,sub2015,sch2015}. 
 For the estimation of the B mode noise, we chose the noise per pixel $\omega_p^{-1/2}=2.8 \mu K$ arcmin corresponding to the LiteBIRD-like experiment \cite{hazu2019}.
From the behavior of the signals which did not increase appreciably for $l\gtrsim 10$ while the noise increased, we could infer that the cumulative signal to noise would not increase appreciably for $l$ far above 10 as verified by this figure.
For $\beta=10^{-4}$ and $10^{-2}$, for our choice of parameters, the noise is dominated by that of the B modes because $C_l^{BB}<N_L^{BB}$ while $C^{2121}_l/N_l^{2121}\sim 5$ for $l\lesssim 100$ (the 21cm observation is hence cosmic variance limited in this example). $C_l^{BB}$ can exceed $N_l^{BB}$ for $\beta=1$, and S/N approaches the cosmic variance limited estimation as shown in this plot. The small difference between the cosmic variance limited $S/N$ and $S/N$ with $\beta=1$ is due to the 21cm noises rather than the B mode ones ($C_l^{BB}/N_l^{BB}\gg C_l^{2121}/N_l^{2121}>1$ for $\beta=1$). 
The sample variance limited $S/N$ is insensitive to the value of $\beta$ in our analysis because both the signal and noise simply scale as $\beta$ in the sample variance limit as can be seen from the Eq. \ref{snformula}. Even with this noiseless limit, the cumulative $S/N$  does not exceed unity and it would be unlikely to be able to detect the 21cm-B mode cross-correlation for our particular model/parameters under discussions. We however note that the actual signals are heavily model dependent and it would be worth exploring other scenarios such as the those involving different redshift dependence and different reionization models. For instance, an earlier reionization scenario can be advantageous for a larger signal because of a larger optical depth and a larger bias at a higher redshift. 
As an illustration, Fig. \ref{snratioplotright} shows the 21cm-B mode cross-correlation signals along with the error (Eq. \ref{errorbar}) for another reionization history where the ionization fraction becomes a half at $z_{rei}=15$ with $\Delta z_{rei}=1$ (even though such a high redshift reionization scenario is disfavored by the current data \cite{ag2016b}). We here chose such $z_{rei},\Delta z_{rei}$ as the inputs of our tanh reionization model (which results in the optical depth $\sim 0.14$). The signal indeed increases for such a higher redshift reionization scenario as expected, but the signal to noise would be still too small for the detection even though the sample variance limited measurements can realize $S/N>1$ and could conceivably detect the signals.

Even though the predicted cross-correlation signal is heavily model dependent, unless the dark energy scalar field evolution changes significantly or the reionization model modifications can lead to the signal enhancement by well more than an order of magnitude compared with our toy model for a realistic value of $\beta$ smaller than $10^{-2}$ \cite{liu2006}, we find that it is unlikely to detect the cross-correlation between the 21cm and the birefringence-induced B mode fluctuations.

\section{Discussions/Conclusions}
\label{discon}
\begin{figure}

                  \includegraphics[width=1.0\textwidth]{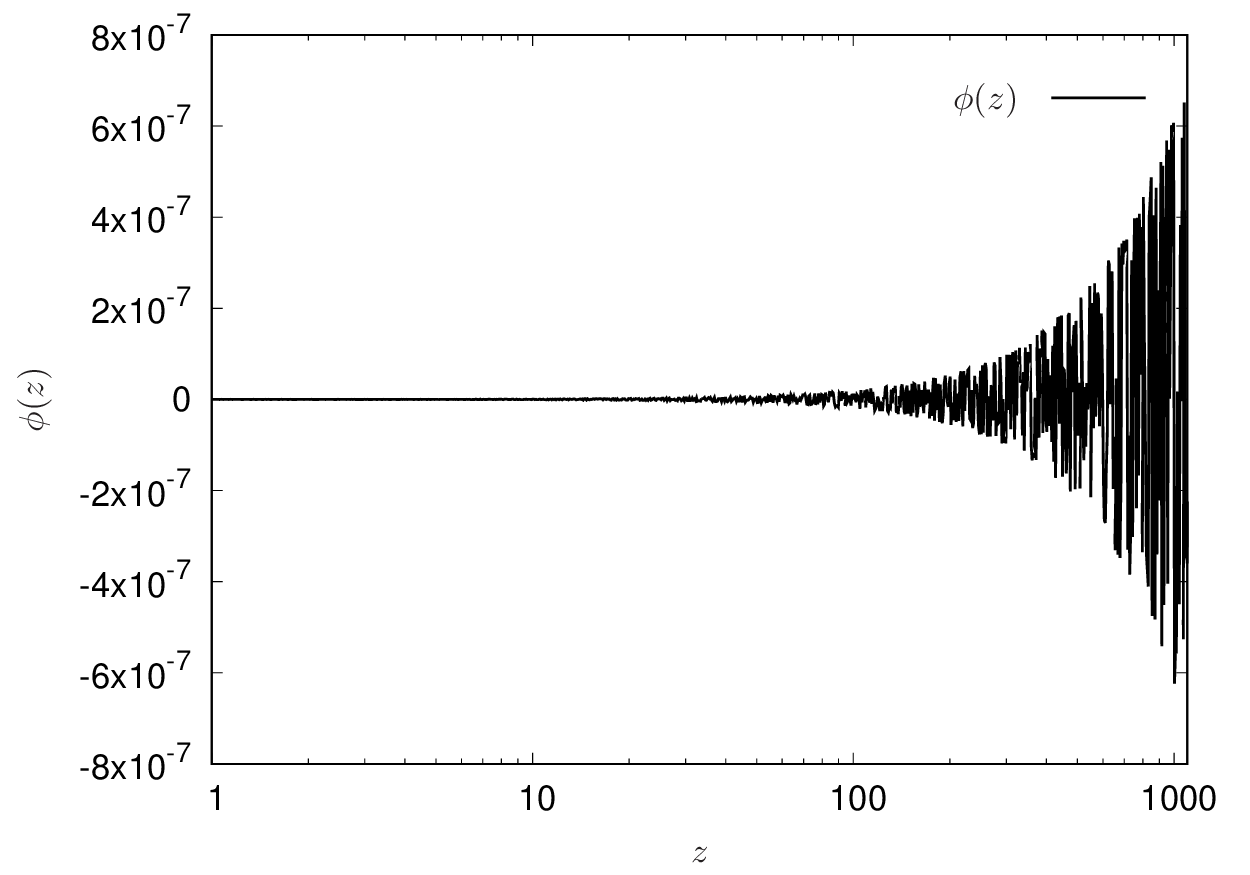}  
  \caption{The evolution of the axion-like scalar field $\phi$ in units of the reduced Plank scale for $m=10^{-22}eV, \beta=(10^{13}GeV)^{-1}$. 
  }
\label{axioncellleft}
\end{figure}

\begin{figure}
  
      \includegraphics[width=1.0\textwidth]{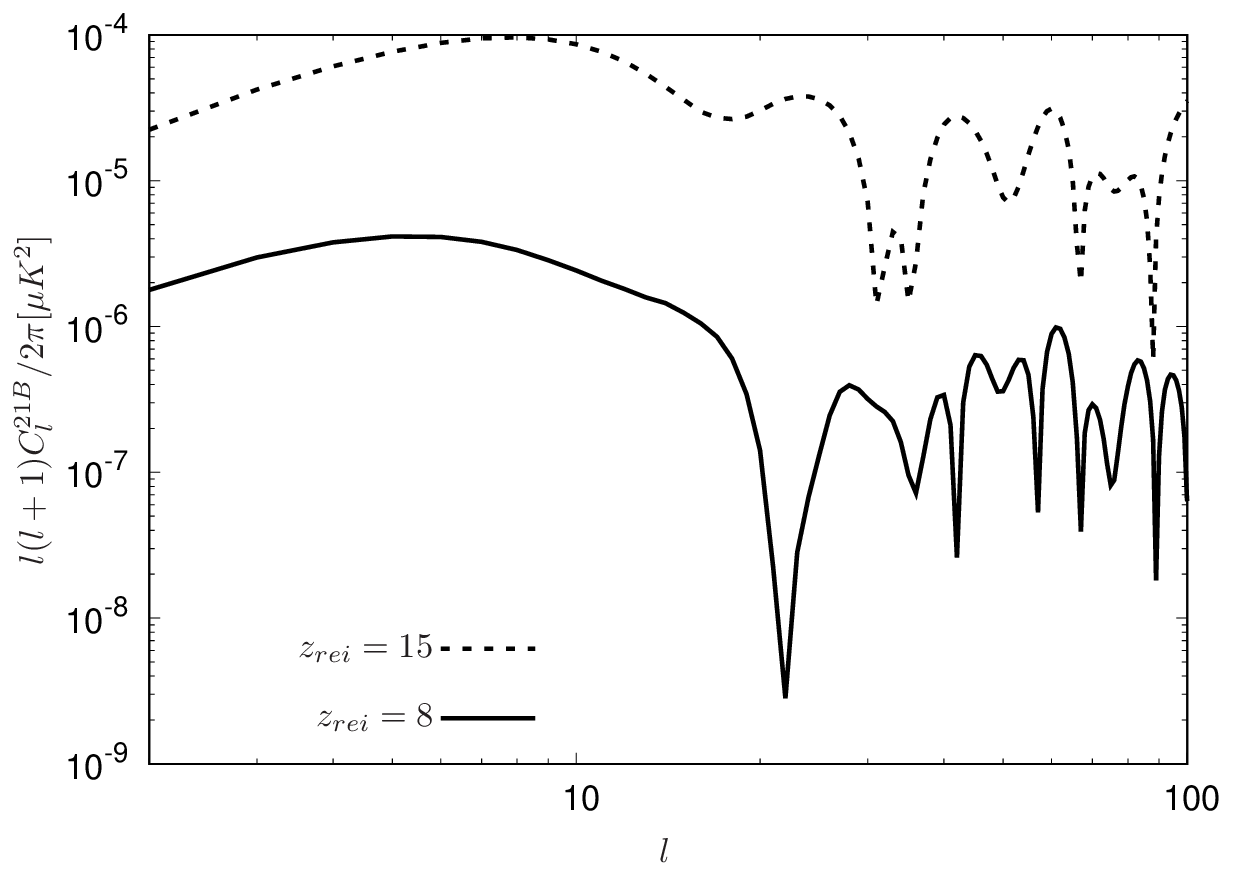} 
  \caption{
    $C_l^{21B}$ (without oscillation cancellation effects) for two reionization scenarios ($z_{rei}=z_{obs}=8$ and $z_{rei}=z_{obs}=15$). $m=10^{-22}eV, \beta=(10^{13}GeV)^{-1}$.}
\label{axioncellright}
\end{figure}

Before the conclusion, let us briefly discuss another common example, the axion-like scalar field responsible for the current dark matter density, which can also lead to the rotation of the photon linear polarization plane. We consider a simple toy potential for this purpose
\ba
V(\phi)=\frac{1}{2} m^2 \phi^2
\ea
The axion-like scalar field evolution is illustrated in Fig. \ref{axioncellleft}. The amplitude was adjusted to match the current dark matter density and we chose $m=10^{-22}eV$ which is motivated to resolve the small scale structure problems (and the mass smaller than this value is in tension with the observations because of a larger de Broglie wavelength within which the structure is suppressed) \cite{hu2000c, hui2016}. We also choose for concreteness $\beta \sim (10^{13}GeV)^{-1}$ which corresponds to the current upper bound from the CMB data for $m=10^{-22}eV$ \cite{agh2018a,fedd2019,sigl2018fba}. 
In contrast to the dark energy example involving the slowly varying background field, the scalar field oscillates rapidly (e.g. compared with Hubble time scale) and can behave as the pressure-less cold dark matter. The birefringence effect can be canceled among the positive and negative rotation angles due to the oscillations which can reduce our desired signals \cite{fedd2019}. Another obstacle to result in the small 21cm-B mode cross-correlation signal of our interest is the small scalar field amplitude around the reionization epoch. While the dark matter scalar field amplitude is large at an earlier epoch, the field amplitude decreases as $(1+z)^{3/2}$ and consequently the rotation angle around the reionization epoch of our interest is too small for the possible detection of the 21cm-B mode cross-correlation signals. This is illustrated in Fig. \ref{axioncellright}  for two reionization histories with the reionization epochs of $z_{rei}=8$ and $z_{rei}=15$ discussed in the last section. These $C_l^{21B}$'s were calculated, to ease the numerical computations and to obtain the conservative estimations, by using the time-dependent field amplitude without oscillations. This $C_l^{21B}$ is certainly an overestimate of the signal because of ignoring the significant cancellation effects from the rapid oscillations, and, yet, suffices for our purpose of demonstrating that the signal is too small for the detection. Even these overestimated signals are indeed smaller than what we found for the dark energy model with $\beta=0.01$ in the last section which was unlikely to be detectable. We hence find that the cross-correlation signal would be too small to be detectable with the forthcoming experiments for such a simple dark matter scalar field model.
\\

With the unprecedented progress in the radio and CMB experiments, there is a growing interest in the 21cm and B mode observations.
Considering the experimental challenges in detecting these signals, verifying the detection and interpretation of those signals by measuring the cross-correlations among different observables would be crucial for our further confidence that the signals are indeed of the cosmological origin we are interested in. 
Motivated by those intensive studies on the 21cm cross-correlation measurements, we studied the 21cm-B mode cross-correlations to probe the cosmological birefringence as an astrophysical test of the fundamental physics.
We have discussed the 21cm-B mode cross-correlation signals in existence of an evolving scalar field using the concrete toy models, the quintessence-like dark energy model and ultra-light axion-like dark matter model. For such simple models/parameters, we found that the cross-correlation signals are too small for the detection. We on the other hand demonstrated through those examples that the signals are heavily dependent on the reionization scenarios and scalar field dynamics, and the further investigation beyond our first attempt on the 21cm-B mode cross-correlation study would be warranted.
The cross correlations also benefit from the removal of the foreground contamination thanks to the different noise dependence on each observable \cite{mina2019,yoshi2018,li2016,mor2012}.
While our paper studied only the homogeneous scalar field background, the anisotropic rotation angles could also be worth exploring and we leave the non-uniform cosmological birefringence and its correlation with the 21cm signals for our future work.
\\
\\
We would like to thank S. Lee, K. Ng, M. Pospelov and J. Silk for the useful discussions. This work was supported by Institute for Basic Science (IBS-R018-D1), Grants-in-Aid for Scientific Research
from JSPS (16J05446, 15K17646, JP15H05890, JP16H01543) and Taiwan Ministry of Science and Technology (104-2112-M-032 -007).


\end{document}